\documentclass[sigconf, authorversion]{acmart}

\usepackage{geometry}
\usepackage{array}
\usepackage{graphicx}
\usepackage{comment}
\usepackage{enumitem}
\usepackage{capt-of}
\usepackage{pgfplots}
\usepackage{caption}
\usepackage{multirow}
\usepackage{float}
\usepackage[utf8]{inputenc}
\usepackage{booktabs}
\usepackage{makecell}
\usepackage{ragged2e}
\usepackage{lipsum}
\usepackage{booktabs}

\AtBeginDocument{%
  \providecommand\BibTeX{{%
    \normalfont B\kern-0.5em{\scshape i\kern-0.25em b}\kern-0.8em\TeX}}}

\setcopyright{none}
\copyrightyear{2024}
\acmYear{2024}
\acmDOI{XXXXXXX.XXXXXXX}

\copyrightyear{2024}
\acmYear{2024}
\setcopyright{rightsretained}
\acmConference[FAccT '24]{The 2024 ACM Conference on Fairness,
Accountability, and Transparency}{June 3--6, 2024}{Rio de Janeiro, Brazil}
\acmBooktitle{The 2024 ACM Conference on Fairness, Accountability, and
Transparency (FAccT '24), June 3--6, 2024, Rio de Janeiro,
Brazil}\acmDOI{10.1145/3630106.3659045}
\acmISBN{979-8-4007-0450-5/24/06}

\makeatletter
\gdef\@copyrightpermission{
  \begin{minipage}{0.3\columnwidth}
   \href{https://creativecommons.org/licenses/by-nc-nd/4.0/}{\includegraphics[width=0.90\textwidth]{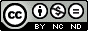}}
  \end{minipage}\hfill
  \begin{minipage}{0.7\columnwidth}
   \href{https://creativecommons.org/licenses/by-nc-nd/4.0/}{This work is licensed under a Creative Commons Attribution-NonCommercial-NoDerivs International 4.0 License.}
  \end{minipage}
  \vspace{5pt}
}
\makeatother

\begin{document}

\title{Visions of a Discipline: Analyzing Introductory AI Courses on YouTube}

\author{Severin Engelmann}
\email{severin.engelmann@cornell.edu}
\orcid{0000-0001-8368-4484}
\affiliation{
  \institution{Cornell Tech, Digital Life Initiative}
  \city{New York City, NY}
  \country{USA}
  }

  \author{Madiha Zahrah Choksi}
\email{mc2376@cornell.edu}
\orcid{0009-0008-4752-7164}
\affiliation{
  \institution{Cornell Tech, Digital Life Initiative}
  \city{New York City, NY}
  \country{USA}
  }

  \author{Angelina Wang}
\email{angelina.wang@princeton.edu}
\orcid{0000-0001-9140-3523}
\affiliation{
  \institution{Princeton University, Computer Science Department}
  \city{Princeton, NJ}
  \country{USA}
  }

  \author{Casey Fiesler}
\email{casey.fiesler@colorado.edu}
\orcid{0000-0002-8743-4201}
\affiliation{
  \institution{University of Colorado Boulder, Information Science Department}
  \city{Boulder, CO}
  \country{USA}
  }

\renewcommand{\shortauthors}{Engelmann, Choksi, Wang \& Fiesler}

\begin{abstract}
Education plays an indispensable role in fostering societal well-being and is widely regarded as one of the most influential factors in shaping the future of generations to come. As artificial intelligence (AI) becomes more deeply integrated into our daily lives and the workforce, educational institutions at all levels are directing their focus on resources that cater to AI education. Yet, informal education, including online learning on social media platforms like YouTube, plays an increasingly significant role for both students and the general public. Offering greater accessibility compared to formal education, millions of individuals use YouTube for educational resources on AI today. Due to the substantial societal impact of AI, it is crucial for introductory AI courses to meaningfully address the ethical implications associated with AI. Our work investigates the current landscape of introductory AI courses on YouTube, and the potential for introducing ethics in this context. We qualitatively analyze the 20 most watched introductory AI courses on YouTube, coding a total of 92.2 hours of educational content viewed by close to 50 million people. We find that these introductory AI courses do not meaningfully engage with ethical or societal challenges of AI (RQ1). When \textit{defining and framing AI}, introductory AI courses foreground excitement around AI's transformative role in society, over-exaggerate AI's current and future abilities, and anthropomorphize AI (RQ2). In \textit{teaching AI}, we see a widespread reliance on corporate AI tools and frameworks as well as a prioritization on a hands-on approach to learning rather than on conceptual foundations (RQ3). In promoting key \textit{AI practices}, introductory AI courses abstract away entirely the socio-technical nature of AI classification and prediction, for example by favoring data quantity over data quality (RQ4). Given the power of openly available introductory courses to shape enduring beliefs around AI and its field at the onset of a learning journey, we extend our analysis with recommendations that aim to integrate ethical reflections into introductory AI courses. We recommend that introductory AI courses should (1) highlight ethical challenges of AI to present a more balanced perspective, (2) raise ethical issues explicitly relevant to the technical concepts discussed and (3) nurture a sense of accountability in future AI developers. 
\end{abstract}





\maketitle

\section{Introduction}\label{introduction}
Artificial intelligence (AI) has become an enduring fixture in the eye of the general public. Daily media reporting on AI often oscillates between two extremes: powerful AI tools that transform how we live, work, and interact with each other, and, on the other hand, doomsday scenarios as a result of AI development ~\cite{bunz2022ai, nguyen2022news}. Recent major regulatory frameworks on AI also tend to define AI in balancing terms around benefits and risks ~\cite{biden2023executive,madiega2021artificial}. Beyond media and policy, \textit{education} is a key contributor to society's knowledge on AI. Education effectively defines and fosters core values around AI, and influences how society perceives AI as an academic discipline. The recent advances in machine learning (ML), deep learning (DL), and generative AI underscore the growing importance of AI education in equipping future AI developers with an acute awareness of the potential risks and adverse impacts resulting from transformative AI systems and applications. However, not all learning takes place in a classroom, and many people are seeking out educational resources about AI \textit{online}.

In this work, we explore how popular introductory AI courses on YouTube present AI to a broad audience. These introductory courses offer unparalleled accessibility to basic AI skills, responding to a growing demand for AI competence in government, industry, academia, and civil society ~\cite{alekseeva2021demand}. YouTube's introductory AI courses exemplify informal online education, eliminating the need for enrollment in study programs such as those available on massive open online course platforms (MOOCs)~\cite{haber2014moocs}. On YouTube, educational content creators showcase their coding process by sharing screencasts, guiding viewers through coding problems step by step~\cite{lee2017youtubecomments}. Such video-based tutorials are particularly valued by developers who, more than learners in other disciplines, have accustomed to a culture of online self-didactic learning ~\cite{luchs2023politicalcourses, heuer2021machine}. 

We focus on introductory AI courses as dynamic spaces for learning where educators \textit{introduce} fundamental definitions, concepts, and practices around AI ~\cite{davis2009tools, allen1996power, de2021m}. In introductory courses, beginners learn what problems the discipline considers to be within its own scope. They introduce the problem-solving approaches that hold value and establish criteria for determining when a problem is considered satisfactorily solved. Such foundational teachings embody some of the core values of a discipline ~\cite{sep-education-philosophy}. In contrast to more advanced students, beginners tend not to question the teaching authority of the instructor and reflect less on the perspectives and assumptions presented in the learning materials ~\cite{sep-civic-education}. AI introductory courses have the potential to leave a lasting impact on students' perceptions of AI and its disciplinary field. 

The introductory AI courses we found on YouTube came from reputable universities, professional educational organizations, and education influencers with large followings. Considering their influential position, introductory AI courses provide a valuable opportunity to cultivate ethical reflection regarding core AI practices right from the start of students' learning journeys. While there is a large corpus of research on computer science and AI education in traditional educational settings (i.e., schools and universities) ~\cite{hazzan2020guide, cooper2010teaching, overmars2004teaching}, and how to best embed ethical considerations into these study programs ~\cite{horton2022embeddedethics, grosz2019embeddedethics,fiesler2021integrating}, less attention has been paid to AI education on YouTube. Our research qualitatively investigates introductory AI courses on YouTube to answer the following four research questions:

\vspace{2mm}

\begin{enumerate}[label=\textbf{RQ\arabic*}., align=left, left=0pt, labelwidth=*]

\item \textbf{Do introductory AI courses on YouTube cover ethics-related topics such as fairness, privacy, or accountability?}
\item \textbf{How do introductory AI courses on YouTube define and frame AI?}

\item \textbf{What tools and exercises do
introductory AI courses on YouTube
focus on?}

\item \textbf{What methodological practices and competencies do introductory AI courses promote?}
\end{enumerate}

\vspace{2mm}

We find that introductory AI courses rarely address ethical or societal harms caused by AI technologies (RQ1). Introductory AI courses promote excitement and hype around AI framing its capabilities by reference to various forms of anthropomorphism (RQ2). Introductory AI courses prioritize hands-on coding at the expense of more conceptual understandings that might nurture a broader perspective of AI's situated uses (RQ3). Finally, introductory AI courses downplay the value of \textit{data competence}. They largely neglect essential skills for appropriate data collection, data curation, and data annotation (RQ4). By focusing almost exclusively on model building and performance they also undervalue the role of domain expertise in prediction and classification tasks. In summary, our study provides evidence that introductory-level AI education on YouTube \textit{promotes AI practices that have been identified as significant contributors to AI-related harms.}

Based on these findings, we formulate recommendations for integrating ethical reflections into introductory AI courses on YouTube in the final part of this work. Teaching ethics in the context of digital technologies is challenging as it requires substantial knowledge and expertise in ethics and computer science scholarship. Prior work on computing ethics in higher education points to the importance of both ethics integration (in-situ ethics learning in technical courses, on the concepts relevant to that course) and standalone courses taught by subject matter experts \cite{fiesler2020syllabi}, though even integration requires some knowledge of the topic and presents some barriers for instructors \cite{smith2023incorporating}. With this in mind, our recommendations are aimed at \textit{promoting} an ethical mindset at the beginning of the learning process. Our recommendations do not necessarily require in-depth ethics subject matter expertise. They prescribe to common theories of ethical interventions around \textit{exposure} and challenge dominant views of computing \cite{padiyath2024}. Our recommendations can be integrated into any introductory AI class independent of the disciplinary background of the instructor. Our contribution is complementary to the important progress that has been made in embedding ethics into computer science curricula in the last years ~\cite{horton2022embeddedethics, grosz2019embeddedethics, fiesler2021integrating, fiesler2020syllabi}. With our recommendations, we seek to support instructors in projecting a more balanced picture of AI and help nurture a sense of accountability among future AI developers in building increasingly powerful AI systems.
\section{Background}\label{background}

\subsection{Online learning}

Prior to the COVID-19 pandemic, online distance learning was primarily viewed as a supplement to traditional forms of education ~\cite{bali2014mooc}. It facilitated convenient exploration of new disciplines and enabled learners to sustain knowledge for career-related purposes. During the COVID-19 pandemic when learners around the world could not attend their educational institutions, online distance learning became the de facto mode of education~\cite{kanetaki2022acquiring, yaacob2020acceptance}. Entire curricula were converted into online programs to prevent learning gaps that could have led to significant inequalities in educational outcomes for millions of students worldwide~\cite{tahat2021influence}. 

Shifting to online formats during COVID-19 legitimized an already existing infrastructure of online learning media as educational programs that can complement, extend, and in some cases even replace conventional education~\cite{adedoyin2023covid,sim2021online}. 

Learners around the world participated in these online videos through massive open online courses (MOOCs) which are produced by universities, educational organizations, and companies hosted on commercial online platforms such as Udacity or Coursera~\cite{baturay2015overview, haber2014moocs}. MOOCs are closely related to the YouTube education videos we study, and both are purported to provide an avenue of education that is more accessible than traditional universities. However, it is contested whether and to what extent MOOCs do democratize education~\cite{dillahunt2014democratizing, hansen2015democratizing, rohs2015moocsdisillusion, lambert2020moocsequity}, though work has found that learners who do not have a college degree are more likely to do better on a MOOC than learners with a college degree~\cite{meaney2022demographic, meaney2023promisemoocs}. MOOCs enrollment is often motivated for career building rather than knowledge building~\cite{zhenghao2015whobenefitmoocs}, and they are typically on a far shorter time scale than a university degree. Additionally, while YouTube does not facilitate social learning (e.g., group assignments) to the same extent as some MOOCs, it does enable interaction with education content creators and other learners through questions and answers in video comments~\cite{lange2019informal}. 

Using online resources for advancing competencies is particularly popular in disciplinary cultures that value and reward self-didactic approaches to learning. One of these disciplines is software development in computer science. In 2021, Stack Overflow --- a popular learning and knowledge sharing platform among software developers --- ran a large-scale survey with more than 80,000 software developers across different industries~\cite{stacko}. When asked \textit{How did you learn to code? Select all that apply}, the results were: 59.5\%  ``Other online resources (videos, blogs, etc)'',  40.4\%  ``Online Courses or Certification'' and 30.6\% ``Online Forum.''\footnote{The remaining responses were:  53.6\% ``School'',  51.5\% ``Books \& Physical Media'' and, with less than 20\%: ``Friend or family member'', ``Colleague'', and ``Coding Bootcamp.''} Online coding tutorials are common at every stage of the learning process for computer science students. Four out of the top ten most ``enrolled'' MOOCs of all times are computer science and programming courses~\cite{topmoocs}. Another factor contributing to the popularity of online learning among developers is the consistent demand for new competencies fueled by continuous technological advancements. YouTube is an especially popular learning resource among coders and developers as content creators can share their screen as screencasts to walk viewers through a coding problem ~\cite{macleod2015code}. This allows educators to illustrate specific coding skills, showcase code documentation and debugging, and explain error messages in relation to the source code. In introductory coding courses, educators can demonstrate how to set up an integrated development environment for coding, which can be challenging for beginners and a drawback of text-based coding tutorials. Moreover, education content creators on YouTube can share personal experiences learning specific competences and recommend what they believe are effective learning strategies~\cite{carpenter2023education}.

All of this is to say that, increasingly, online learning takes place on social media platforms such as YouTube~\cite{lausa2021educational} or even TikTok~\cite{garcia2022tiktok} where learners search for and easily access educational content in a familiar interface. 
For comparison, enrollment in MOOCs require account registration and, in many cases, payment after a ``freemium'' period. 
The barrier to upload content on YouTube is lower than for MOOCs platforms. Among other effects, this diversifies the community of education content creators. In addition to universities\footnote{Some universities transfer entire course lectures to YouTube. For example, UC Berkeley uploaded lecture courses in bioengineering and conflict studies to YouTube as early as 2008~\cite{fralinger2009you}.} and professional educational organizations (e.g., bootcamps and certification providers) a plethora of education influencers regularly publish learning materials on YouTube~\cite{portugal2018free}. Some social media education influencers have millions of followers and have become ``mini celebrities'' ~\cite{carpenter2023education}. 

On YouTube, this results in a diverse array of video-based learning formats, encompassing lectures, tutorials, walk-throughs, guides, vlogs, and informational videos ~\cite{lange2019informal}. Education content creators have the freedom to teach as they prefer. Consequently, video lengths vary, ranging from concise tutorials lasting a couple of minutes to more extensive ones that extend over several hours ~\cite{poche2017analyzing, macleod2015code}. The following subsection traces relevant works on the education of AI topics considering learning formats beyond only online learning.

\subsection{AI Education}

In recent years there has been growing attention to best practices for AI education, from college courses~\cite{allen2021toward} to middle school \cite{lee2021developing} to informal public education~\cite{long2021co}. Pedagogical strategies for introducing AI has been a topic of research and discussion for decades, and as early as the 1990s, there was recognition of “the oft-voiced complaint” that introductory AI is a difficult topic to teach~\cite{wollowski2016survey}. A 2004 survey of introductory AI instructors revealed that part of this problem stems from the difficulty in keeping up with the latest research trends, resulting in teaching outdated content~\cite{harris2004pedagogy}. This survey also did not reveal ethics as a component of these instructors’ courses, and similarly, a 2021 literature review of AI education had very little mention of ethics~\cite{allen2021toward}. However, a recent synthesis of AI education literature towards defining components of AI literacy lists ethics as a core competency, describing many of the core commitments of the FAccT community such as fairness and transparency~\cite{long2021co}. As Ng et al. pointed out in their review of AI literacy literature specifically, knowing how to use AI for future careers is only one relevant aspect, and an understanding of ethical issues is essential for ensuring AI for social good~\cite{ng2021conceptualizing}.

Ethics interventions in computing education have been growing in number, though it is worth noting that ``ethics'' is often used as shorthand for related concepts such as responsibility, social impact, and justice~\cite{brown2023teaching}. A recent review of ethics interventions in computing education also revealed a number of shared goals, including exposure (which assumes a lack of existing awareness of relevant ethical issues, an assumption that is supported by prior work), positioning students as practitioners who will be accountable for their decisions, and challenging dominant or common views of technology~\cite{padiyath2024}. There are also arguments for including ethical concepts explicitly in introductory computing courses, including that it helps to create a "culture" around ethics as a component of technical practice, and it also ensures that even students who take only one lower-division course are exposed to ethics ~\cite{fiesler2021integrating}. One intervention in an introductory AI course that introduced ``human impact'' as a necessary constraint in an algorithm assignment had similar goals, and found that this strategy had the potential to deepen technical rigor while also supporting instructor pedagogical content knowledge~\cite{brown2022shortest}.

Though in recent years discussion of these types of integrative ethics interventions have increased~\cite{goetze2023integrating}, there are barriers to instructors in doing so~\cite{smith2023incorporating} and even initial stages such as identifying an appropriate ethical context for assignments can be a challenge~\cite{brown2023designing}. A 2018 review of syllabi for AI, machine learning, and data science courses revealed that only 12\% of such courses included any (generously defined) mention of ethics-related concepts~\cite{saltz2019integratingethics}. However, when raised, common relevant topics in AI courses include bias, privacy, and unintended consequences~\cite{garrett2020timeallows}.

Despite the potential benefits for teaching ethical concepts in-situ alongside relevant AI technical topics, and of teaching applied ethics in computer science departments, it is critical to keep in mind as articulated by~\citet{raji2021you} that the ``myth of the ethical unicorn'' in computer science accompanies a devaluing of social science and humanities. The ``exposure'' theory may be a good starting place (particularly coming from nothing), but AI education also needs to include engagement with humanistic fields and learn from deep subject matter experts outside of technical computer science. One strategy beyond pointing students to separate courses is for instructors from different disciplines to work collaboratively to create and deliver educational content--for example, philosophers creating computing ethics modules~\cite{grosz2019embedded} or NLP and accessibility experts working together to teach students about AI fairness~\cite{dobesh2023towards}. However, it is important that in such interdisciplinary approaches technical skills are not privileged and valued above ethical skills~\cite{raji2021you, goetze2023integrating}.

Additionally, beyond formal education, recent projects aimed at the general public have also presented the important goal of increasing legibility of AI for non-experts~\cite{hemment2023ai}, emphasizing that even a casual understanding of AI (understanding how a system works as opposed to how to build one) is growing in importance~\cite{long2021co}. Our work here was inspired in part by this acknowledgment of the importance of informal AI education, as well as the potential benefits for including ethics as part of introductory courses.

\section{Methods}\label{methods}
\subsection{Sampling Introductory AI You-Tube Videos}

We sourced English-language introductory AI/ML videos on YouTube. We approached our search for introductory AI courses by envisioning how an individual eager to start learning AI would likely try to find relevant educational resources on YouTube. We employed a keyword-based search on YouTube for our selection criteria as educational tutorials tend to specify ``intro to AI'' (or equivalent descriptives) in their title. For our queries, we used the following keywords: \textit{introduction to AI}, \textit{introduction to artificial intelligence}, \textit{introduction to ML}, \textit{introduction to machine learning}, \textit{introduction to DL}, \textit{introduction to deep learning}, \textit{beginner course AI}, \textit{beginner course artificial intelligence}, \textit{beginner course ML}, \textit{beginner course machine learning}, \textit{beginner course DL}, \textit{beginner course deep learning}. For our sample of introductory AI videos on YouTube, we followed the search procedures outlined in ~\cite{heuer2021machine}. This included utilizing a computer that had never been used before for the purpose of searching introductory AI courses on YouTube. To minimize any potential effects of personalization, the search was performed in private browser mode and after deleting cookies. For each query, we ranked the results by most viewed and started selecting videos in accordance with the following criteria. We specified that videos had to cover at least one technical concept, such as basic AI algorithms, parts of the machine learning pipeline, or relevant concepts (e.g., a perceptron). This requirement meant exclusion of videos that introduced AI in a non-educational context, for example, videos related to best learning approaches (e.g., ``The best way to learn AI in 2024'') or comparisons like ``AI vs. ML.'' Moreover, videos had to be at least 15 minutes. Our final dataset includes the 20 most viewed introductory AI courses on YouTube. Considering the extensive collection of search results, we stopped sampling when the next most viewed video had less than 1 million views, which was at N=20.\footnote{Note that YouTube's interface does not show search results in page tabs but in continuous scroll.} We present key information of the videos in our final sample in Table ~\ref{tab:definite_sample}. 

\subsection{Coding Process}

Our methodology combines both deductive and inductive approaches to qualitative coding. After gathering our final corpus of introductory AI courses, we generated full transcripts for each introductory AI course using \textit{youtubetranscript.com}, a free tool that uses Merlin AI to transcribe YouTube videos. 
Coding our sample of YouTube videos posed unique challenges due to substantial variations in course length, ranging from 15 minutes to over 10 hours. Consequently, the topics covered in these courses exhibited significant diversity, with shorter courses focusing on specific technical concepts, while longer courses covered a broader array of AI concepts and coding examples. Another challenge in our coding process was the diverse structure of courses, as instructors had complete agency in organizing their content. Some courses began with an introduction to AI, while others immediately delved into coding examples. We thus opted to code colloquial sentence structures, dealing with highly unstructured data compared to more prompted interview studies. We employed Taguette,\footnote{\url{https://www.taguette.org/}} a web-hosted collaborative and open-source tool for qualitative coding. 



We followed best practices for coding qualitative data~\cite{mcdonald2019reliability, krippendorff2018content, bengtsson2016plan}. Applying researcher triangulation~\cite{patton1999enhancing}, three interdisciplinary researchers separately assessed V1 (university) to inductively identify first emerging patterns and trends. Researchers then met to present and discuss their initial codes. This resulted in a consolidation of a first set of codes (e.g., codes \textbf{\textit{High-level AI definition} and \textbf{Excitement around AI}}). This process was repeated for V2 (influencer) and V3 (online education organization), with discussions on and iterations to the set of codes after each video. Researchers met three times to discuss the coding scheme and made adjustments given the structural variability of introductory AI videos in our sample (e.g., length, content publisher). The final codebook can be organized around three major themes: definitions and framing of AI, teaching resources and styles, and practicing AI. This organization is based on high-level themes reflecting the descriptive and evaluative observations about specific elements within the courses. We present the final codebook with 21 codes in Table \ref{tab:codeBook1} in the Appendix. Ongoing discussions among the research team further shaped the codebook, with occasional decisions to merge codes, especially when addressing general approaches to data and data collection practices. For example, researchers removed the code \textbf{\textit{Data Privacy}} which was collectively coded 3 times within 20 videos. It is crucial to note that instances of disagreement during these discussions are documented for transparency. The remaining videos in our sample were coded deductively using the established codebook. Taguette allows coders to apply multiple codes to a single textual unit, accommodating the nuanced nature of the content.

After the third discussion session (i.e., post coding V3), once consensus on a codebook was reached, three of the authors independently coded V4 (influencer, approximately 4 hours) to calculate Inter-Rater Reliability (IRR) scores. The obtained IRR measures for this coding task reflect strong agreement among the research team. Specifically, the Fleiss Kappa coefficient, which assesses agreement beyond chance among multiple raters was calculated at 0.873. Additionally, Cohen's kappa, a statistic that considers agreement between two raters while accounting for chance calculated at 0.876. Krippendorff's alpha was calculated at 0.877. These multiple reliability indices affirm high agreement and consistency in our qualitative coding. 

It is also worth noting the perspective presented by McDonald et al. that finding informal consensus through discussion meetings suffices to communicate reliability, particularly in the context of highly unstructured data and multi-label coding ~\cite{mcdonald2019reliability}. This perspective aligns with our methodology, emphasizing the importance of discussion and consensus-building in ensuring the reliability of our qualitative analysis. Finally, we acknowledge our positionality relative to the study and provide ethical considerations in Section \ref{position}, which is presented at the very end of this work. 

\begin{table*}[htbp]
    \centering
    \begin{tabular}{cccc}
        \toprule
        \textbf{Video} & \textbf{Content Creator} & \textbf{Publication Year} & \textbf{Technical Concept (rand. ex.)} \\
        \midrule
        V1 & University & 2023 & Perceptron \\
        V2 & Influencer & 2022 & Matrices \\
        V3 & Online education organization & 2018 & Input layer \\
        V4 & Influencer & 2022 & Computing the loss \\
        V5 & Influencer & 2021 & Training and testing dataset \\
        V6 & Online education organization & 2020 & Indexing \\
        V7 & Online education organization & 2019 & Linear regression \\
        V8 & Influencer & 2020 & Markov model \\
        V9 & Influencer & 2020 & Decision tree \\
        V10 & Online education organization & 2019 & Random forest \\
        V11 & University & 2020 & Convex optimization algorithms \\
        V12 & Influencer & 2019 & Batch normalization \\
        V13 & University & 2014 & Neural network \\
        V14 & University & 2017 & Feature engineering \\
        V15 & Influencer & 2020 & Collective intelligence \\
        V16 & University & 2023 & Greedy best-first search \\
        V17 & Influencer & 2022 & Tensors \\
        V18 & Influencer & 2017 & Weights and biases \\
        V19 & Company & 2023 & Discriminative model \\
        V20 & Influencer & 2021 & Sigmoid function \\
        \midrule
        \multicolumn{2}{l}{\textbf{Total views}: 49,217,851} & \multicolumn{2}{l}{\textbf{Total length}: 92:22:07 hours}  \\
        \multicolumn{2}{l}{\textbf{Average views}: 2,460,893} & \multicolumn{2}{l}{\textbf{Average length}: 4:37:06 hours} \\
        \multicolumn{2}{l}{\textbf{Median views}: 1,559,587} & \multicolumn{2}{l}{\textbf{Median length}: 1:11:42 hours} \\
        \bottomrule
    \end{tabular}
    \caption{The final sample of the 20 most watched introductory AI courses on YouTube. The sample comprises three different content creators (universities,  influencers, and online education organizations) and spans the years 2014 - 2023. Randomly selected \textit{\textbf{Technical Concepts}} taught in the courses are displayed in the right column of the table.}
    \label{tab:definite_sample}
\end{table*}

\section{Results}\label{sec_results}

\subsection{RQ1: Do introductory AI courses on YouTube cover ethics-related topics acknowledging that AI can cause harm?}

\subsubsection{Ethics plays no significant role in introductory AI courses on YouTube} 
\label{ethicskeywords}

First, we conducted a keyword search of ethics-related concepts in the transcripts of introductory AI courses. We employ the word stems of the following core ethics-related concepts: \textit{ethics, privacy, fairness, accountability, transparency, responsibility, bias, explainability, discrimination}.\footnote{Our analysis only considers use of these concepts in an ethical or societal context. We exclude counting these words in different contexts. For example, the term ``bias'' appeared frequently in introductory AI courses, however, our manual coding revealed exclusive use of the term in the context of model parameters and weights. Instructors used the term ``fair'' in statements such as: ``We're covering a fair bit of ground today'', which we did not consider for the analysis.} We selected these ethics-related concepts given their prominence in AI ethics education (e.g.,~\cite{garrett2020timeallows}) and in the FAccT literature (e.g.,~\cite{laufer2022four}). In Table \ref{tab:ethicskeywords}, we show the prevalence of these ethics terms across introductory AI courses on YouTube. 

The majority (17) of introductory AI courses in our corpus did not mention or otherwise meaningfully address ``ethics.'' Of the 3 courses that did mention ``ethics'' instructors did not go beyond general statements such as ``people are doing work on AI ethics'' (V12, V15) or ``humans need to behave ethically'' (V11) to support machines solving people's problems. Only V15 mentioned ``privacy,'' ``fairness,'' ``responsibility,'' and ``bias'' when introducing AI (1/20). Apart from V15, a course introduction by an influencer that offered a more in-depth overview of the AI field in general, ethical and societal considerations of AI technology did not play a significant role in introductory AI courses on YouTube. In V10, the instructor emphasized the importance of ``transparency,'' stating that the mathematical notation for linear regression is easy to grasp. In V17, the instructor stated explainability of deep learning models is challenging albeit important for human interpretation. Not a single introductory AI course mentioned or otherwise addressed any issues of ``accountability'' or ``discrimination'' in the context of introducing AI. For introductory AI courses that were part of a lecture series, we searched all remaining lectures for the prevalence of any any ethics-related course sessions in any of the lecture series of introductory AI courses in our sample. Adding to these observations, in our qualitative, line-by-line coding of the video transcripts, we did not identify any implicit or indirect teaching of ethics or ethics-related concepts. 

We assessed whether the content creators of introductory AI courses in our sample had uploaded any courses on AI ethics, which was not the case. Because we considered that perhaps AI ethics is covered in other educational videos on YouTube, we ran another keyword search to source introduction/beginner AI \textit{ethics} courses, applying our previous sampling criteria. Search results indicated a patchy landscape of different topics including video titles such as ``Will AI take my job?'', ``AI ethics consulting'', or ``The truth about AI''. In total, we found seven AI ethics courses with at least 1000 views. Only V2 offered a full introduction to AI ethics. Other courses addressed a specific AI technology (e.g., generative AI), a more concrete problem in AI ethics (e.g., bias and fairness) or a subfield of AI ethics (e.g., responsible AI). In comparison to introductory AI courses, AI ethics courses were significantly less represented on YouTube. We summarize general information on the most watched AI ethics courses on YouTube in Table \ref{tab:ethicscourses}. Notably, these videos have considerably fewer views than the introduction to AI videos in our corpus.

\begin{table*}[htbp]
    \centering
    \renewcommand{\arraystretch}{1.5}
    \resizebox{\textwidth}{!}{
        \begin{tabular}{|c|*{22}{c|}}
            \hline
            \textbf{} & \textbf{V1} & \textbf{V2} & \textbf{V3} & \textbf{V4} & \textbf{V5} & \textbf{V6} & \textbf{V7} & \textbf{V8} & \textbf{V9} & \textbf{V10} & \textbf{V11} & \textbf{V12} & \textbf{V13} & \textbf{V14} & \textbf{V15} & \textbf{V16} & \textbf{V17} & \textbf{V18} & \textbf{V19} & \textbf{V20} & \textbf{Total} & \textbf{In \# videos} \\
            \hline
            \textbf{Ethics} & 0 & 0 & 0 & 0 & 0 & 0 & 0 & 0 & 0 & 0 & \textbf{1} & \textbf{3} & 0 & 0 & \textbf{11} & 0 & 0 & 0 & 0 & 0 & \textbf{15} & \textbf{3/20} \\
            \hline
            \textbf{Privacy} & 0 & 0 & 0 & 0 & 0 & 0 & 0 & 0 & 0 & 0 & 0 & 0 & 0 & 0 & \textbf{4} & 0 & 0 & 0 & 0 & 0 & \textbf{4} & \textbf{1/20} \\
            \hline
            \textbf{Fairness} & 0 & 0 & 0 & 0 & 0 & 0 & 0 & 0 & 0 & 0 & 0 & 0 & 0 & 0 & \textbf{2} & 0 & 0 & 0 & 0 & 0 & \textbf{2} & \textbf{1/20} \\
            \hline
            \textbf{Accountability} & 0 & 0 & 0 & 0 & 0 & 0 & 0 & 0 & 0 & 0 & 0 & 0 & 0 & 0 & 0 & 0 & 0 & 0 & 0 & 0 & \textbf{0} & \textbf{0/20} \\
            \hline
            \textbf{Transparency} & 0 & 0 & 0 & 0 & 0 & 0 & 0 & 0 & 0 & \textbf{2} & 0 & 0 & 0 & 0 & \textbf{1} & 0 & 0 & 0 & 0 & 0 & \textbf{3} & \textbf{2/20} \\
            \hline
            \textbf{Responsibility} & 0 & 0 & 0 & 0 & 0 & 0 & 0 & 0 & 0 & 0 & 0 & 0 & 0 & 0 & \textbf{4} & 0 & 0 & 0 & 0 & 0 & \textbf{4} & \textbf{1/20} \\
            \hline
            \textbf{Bias} & 0 & 0 & 0 & 0 & 0 & 0 & 0 & 0 & 0 & 0 & 0 & 0 & 0 & 0 & \textbf{2} & 0 & 0 & 0 & 0 & 0 & \textbf{2} & \textbf{1/20} \\
            \hline
            \textbf{Explainability} & 0 & 0 & 0 & 0 & 0 & 0 & 0 & 0 & 0 & 0 & 0 & 0 & 0 & 0 & 0 & 0 & \textbf{2} & 0 & 0 & 0 & \textbf{2} & \textbf{1/20} \\
            \hline
            \textbf{Discrimination} & 0 & 0 & 0 & 0 & 0 & 0 & 0 & 0 & 0 & 0 & 0 & 0 & 0 & 0 & 0 & 0 & 0 & 0 & 0 & 0 & \textbf{0} & \textbf{0/20} \\
            \hline
        \end{tabular}
    }
    \caption{Prevalence of ethics concepts in introductory AI courses on YouTube. Numbers in cells indicate the frequency of ethics concepts mentioned in V1 to V20.}
    \label{tab:ethicskeywords}
\end{table*}

\begin{table*}[htbp]
    \centering
    \begin{tabular}{cccc}
        \toprule
        \textbf{Video} & \textbf{Content Creator} & \textbf{Publication Year} & \textbf{Topical Focus}  \\
        \midrule
        V1 & University & 2021 & Bias \& fairness \\
        V2 & Influencer & 2023 & AI ethics \\
        V3 & University & 2023 & AI ethics bias \\
        V4 & Online education organization & 2023 & Risks of generative AI \\
        V5 & Online education organization & 2018 & Ethic of AI warfare \\
        V6 & Online education organization & 2022 & Responsible AI \\
        V7 & Company & 2020 & Ethics of AI \\
        \midrule
        \multicolumn{2}{l}{\textbf{Total views}: 208,620} & \multicolumn{1}{l}{\textbf{Total length}: 6:32:59 hours} \\
        \multicolumn{2}{l}{\textbf{Average views}: 29,802} & \multicolumn{1}{l}{\textbf{Average length}: 0:56:08 hours} \\
         \multicolumn{2}{l}{\textbf{Median views}: 36,919} & \multicolumn{2}{l}{\textbf{Median length}: 0:48:06 hours} \\
        \bottomrule
    \end{tabular}
    \caption{Overview of courses on YouTube that offered educational content on AI ethics. Only V2 presented a comprehensive overview of AI ethics. Remaining videos presented specific content such as ethics of AI warfare.}
    \label{tab:ethicscourses}
\end{table*}

\subsection{RQ2: How do introductory AI courses on YouTube define and frame AI?}
We find two relevant axes in how videos introduce AI initially: with excitement and hype around the topic and through a formative set of definitions and categorizations.
    
\subsubsection{Excitement and hype: Introductory AI courses draw an entirely positive image of AI and its field}
\label{excitement}


In 18 introductory AI courses, instructors promoted an overly positive impression of AI and its adjacent field (codes \textbf{\textit{Timeliness \& Urgency of AI}} and \textbf{\textit{Excitement \& Hype around AI})}. During the initial phase of the course, instructors typically claimed that the field of AI had arrived at a particularly important point in its development in order to create a sense of urgency around entering this field \textit{now}. Instructors framed recent AI advances to be ``\textit{particularly rapid}'' (V11) and ``\textit{accelerating faster than ever before}'' (V1). In three more recently published courses (2022 \& 2023), instructors started the lesson underlining the fast development of AI by referring to the widespread use of generative AI (V1, V19) and ChatGPT (V2). Explaining the timeliness and urgency of AI development, course instructors typically pointed to one, two, or all of the following reasons: (1) abundance of data, (2) access to fast compute, and (3) maturity of large-scale algorithms. We present illustrative textual examples of our code \textbf{\textit{Timeliness \& urgency of AI}} in Table \ref{tab:urgency}.

\begin{table*}[htbp]
    \centering
    \renewcommand{\arraystretch}{1.5}
    \resizebox{\textwidth}{!}{
        \begin{tabular}{>{\RaggedRight}m{0.9\textwidth}}
            \toprule
            \textbf{Timeliness \& Urgency of AI} \\
            \midrule
            ``Many incredible successes and a lot of problems that even just a decade ago we thought were not really even solvable in the near future now we're solving with deep learning with incredible ease.'' (V1) \\
            ``It's 2023. Tools like chat GPT exist and all of a sudden a lot of you want to dive into machine learning and artificial intelligence.'' (V2) \\
            ``Probably one of the most exciting advancements that we're in the middle of experiencing as humans.'' (V3) \\
            ``Right now it's the most important technology in today's world.'' (V7) \\
             ``There’s one subset of machine learning that’s really hot right now because it’s just advancing very rapidly, which is deep learning.'' (V11) \\
            \bottomrule
        \end{tabular}
    }
    \caption{Illustrative examples of code \textit{\textbf{Timeliness \& Urgency:}}
    During the first phase of the introductory course, and throughout, instructors framed AI as one of the most important technologies, solving challenges previously deemed unsolvable, and praised recent advances in particular.}
    \label{tab:urgency}
\end{table*}

Further enhancing the timeliness of AI progress, instructors created hype around AI by placing emphasis on AI's power and impact across different disciplines, fields and industries; by encouraging the audience to learn AI as these skills could lead to a highly profitable career; and by portraying AI and the ``\textit{AI community}'' as ``\textit{exciting}'' and ``\textit{fun}.'' \newline \textbf{Instructors framed AI to be among the most powerful technologies today}. The audience was told that AI skills allowed them to participate in the development of ``\textit{self-driving cars}'' (V7), the ``\textit{future of autonomy}'' (V1), or ``\textit{self-aware machines}'' (V3). In V15, the instructor stated that AI contributes to ``\textit{humanity at large}'' and, in V11, introduced AI as a technology that can ``\textit{make our democracy run better rather than make it run worse}''. In V12, the audience was told that ``\textit{AI began with the ancient wish to forge the gods}''. Instructors emphasized AI to be a universal technology able to solve problems across a wide range of application fields. \textbf{Instructors told the audience that learning AI would likely lead to high-paying jobs}: connecting to statements around AI's universal applicability, courses promoted AI skills to be in demand across many different industries and thus economically lucrative (``\textit{I'm sure you all agree that machine learning is the hottest trend in today's market}'' (V10)). The importance of acquiring AI skills was justified by reference to a more general development toward automation industry processes (``\textit{Everyone is going for automation}'' (V6)). \textbf{Instructors told the audience that AI  was an exciting field:} Instructors told the audience that AI was ``\textit{exciting}''\footnote{Based on a keyword search and subsequent manual analysis, we found that courses using the word ``\textit{exciting}'' in relation to AI or the field of AI most often were V17 (66 times), V15 (59 times), V12 (21 times), V11 (11 times) and V1 (7 times).} and that they could join an ``\textit{enthusiastic}'' AI community (V1). We show illustrative examples of our code \textit{\textbf{Excitement \& Hype around AI}} in Table \ref{tab:excitement}.  

In contrast, a minority of introductory AI courses (7/20) attempted to balance out the excitement and hype around AI at some point throughout the session (code \textbf{\textit{Reducing hype around AI}}). However, this counter-balance largely came in the form of acknowledging the difficulty or drudgery of AI work rather than its impact on the world. For example, some courses indicated AI/ML development can be frustrating and challenging work; for example, in V6, the instructor stated: ``\textit{70 to 80\% of the time in a ML project you're not really sitting and programming you're just trying to understand your data and that's true I mean I’m not making this up.}''  V2 and V17 (both influencers) also offered alternatives to the overly positive introduction of AI within the first phase of the course: ``\textit{If you get into this field most of your work is not going to be training a model, coming up with a model architecture and doing all of that fun stuff you might see in YouTube videos or hear about in articles...this is not as glorious and as glamorous as a field as you might make it out to be.}'' (V2). In only one course (V17) did the instructors offer an explicit counter to the hype that began most videos,  pointing out that ``\textit{...although machine learning is very powerful and very fun and very exciting, it doesn't mean that you should always use it...Machine learning isn't a solve all for everything}.'' 

\begin{table*}[htbp]
    \centering
    \renewcommand{\arraystretch}{1.5}
    \resizebox{\textwidth}{!}{
        \begin{tabular}{>{\RaggedRight}m{0.9\textwidth}}
            \toprule
        \textbf{Excitement \& Hype around AI} \\
        \midrule
        ``Go, go transform healthcare or go transform transportation or go build a self-driving car. Um, and do all of these things that, um, after this class, I think you'll be able to do.'' (V11) \\
      ``PyTorch 1.0 and TensorFlow 2.0. these really solidified, exciting, powerful ecosystems of tools that enable you to do very, to do a lot with very little effort. The sky is the limit, thanks to the tooling.'' (V12) \\
        ``We can train autonomous vehicles entirely in simulation and deploy them on full-scale vehicles in the real world seamlessly.'' (V1) \\
        ``The average annual salary for a machine learning engineer, which is over \$134,000. And there are also a lot of job openings...''(V7)\\
        ``With python being so easy to learn and having so powerful capabilities it seems to be the dominant choice for new learners and industry professionals alike...Who knows? You might be the next industry expert.'' (V6)\\
        \bottomrule
       \end{tabular}
    }
    \caption{Illustrative examples of code \textit{\textbf{Excitement \& Hype around AI}}: During the first phase of the course, instructors framed AI skills as powerful competencies in high demand across different high-paying industries and with near-universal application possibilities.}
    \label{tab:excitement}
\end{table*}

\subsubsection{Framing AI: High-level definitions, AI/ML examples, and AI abilities (present and future)} \label{definitions}

In 19 introductory AI courses, instructors presented a general, high-level definition of AI or AI sub-fields such as machine learning and deep learning (code \textbf{\textit{High-level definition of AI})}. These definitions typically framed AI as a ``\textit{universal}'' technique that enables the ``\textit{automatic extraction of information from data}'' without a human having to ``\textit{explicitly}'' program a machine (or a computer) to do so (e.g., V3, V5, V7, V11, V14). We found various different AI definitions across our corpus of introductory AI courses ranging from pattern extraction from data over learning a mapping function to automating human decision-making to machines that can think and act like humans. Here, instructors often highlighted the need for AI algorithms to generalize to new data (code \textbf{\textit{Generalizability}}).\footnote{For example, in V1, the instructor explained: ``\textit{We want to build models that can learn representations from our training data that can still generalize even when we show them brand new unseen pieces of test data.}''} Additionally, instructors defined AI taxonomically. While some described the delineations of 1) supervised, 2) unsupervised and 3) reinforcement learning (and in rare cases, semi-supervised learning), others chose more capability-based delineations. For example, 1) reactive, 2) limited memory, 3) theory of mind, 4) self-aware (V4), or even 1) artificial narrow intelligence, 2) artificial general intelligence, 3) artificial super intelligence (V7). 

Similarly, we found various claims about AI's current abilities: from performing ``\textit{complex calculations}'' (V16) to ``\textit{problem solving}'' (V9) and  ``\textit{processing millions of documents in 10 seconds}'' (V7) to ``\textit{AI building software}'' (V1) and ``\textit{robots that can serve coffee}'' (V3) (code \textbf{\textit{AI ability (now)}}). Claims about future abilities of AI typically made references to ``\textit{sentient and conscious}'' machines (V3), ``\textit{artificial super intelligence}'' (V7), ``\textit{general purpose artificial intelligence}'' (V12) or ``\textit{level 4 AI}'' (V15) that does not require any ``\textit{human supervisor}'' (code \textbf{\textit{AI potential in future}}). When defining AI and describing its abilities, we noticed that instructors often attributed human characteristics to AI (code \textit{\textbf{Anthropomorphizing AI}}). For example, when a model was not trained it was compared to a baby that has no information about the world (V1) while a trained model was compared to ``\textit{an intelligent person}'' (V9). In Table \ref{tab:anthro}, we show representative statements made by instructors that attributed human capacities to AI (``\textit{thinking}'', ``\textit{reasoning}'', ``\textit{learning}'' like humans) or that compared AI abilities to human abilities. The tendency to anthropomorphize AI was particularly prevalent in introductory AI courses: 17 out of 20 courses attributed human characteristics to AI or compared AI to human performance. 

We found that only a minority of introductory AI courses (7/20) addressed limitations of AI (code \textbf{\textit{AI limitations}}). In V1, the instructor outlined the computational cost of computing gradient descent in a dataset while, in V12, the instructor stated AI to be ``\textit{extremely smart at a particular task but not...able to transfer}'' to another task. V15 told the audience that AI ``\textit{doesn't understand the fundamentals of the concepts that are being reasoned about}'' and, in V19, the instructors outlined that ``\textit{hallucinations can be a problem for transformers because they can make the output text difficult to understand. They can also make the model more likely to generate incorrect or misleading information.}'' Overall, introductory AI courses did not consistently cover fundamental limitations of AI. 

Moving from a general, high-level definition of AI to a specific context application or problem task, instructors offered the audience a first ``\textit{real-life}'' example of an AI application (code \textbf{\textit{AI/ML example}}, 17/20).\footnote{In some introductory classes, these AI/ML examples also served as prediction/classification coding examples, see AI/ML use cases in Table \ref{tab:codingex} in the Appendix.} We categorized all AI/ML examples instructors presented in introductory AI courses in Table \ref{tab:aiexamples} in the Appendix. Examples from the field of computer vision were presented most frequently. The most common AI/ML examples were differentiating images of cats versus images of dogs (8/20), scene recognition in self-driving cars (8/20) and face recognition (7/20). Introductory AI courses presented a plethora of different AI/ML example applications ranging from generative AI (a deepfake of former US President Obama welcoming the class, (V1)), smart home (intelligent heating, (V3)) to medical predictions (diabetes, cancer and COVID-19, (V4, V5)), gaming (boat racing, (V12)) and recommender systems (Netflix, Google search engine, (V7)).

\begin{table*}[htbp]
    \centering
    \renewcommand{\arraystretch}{1.5}
    \resizebox{\textwidth}{!}{
        \begin{tabular}{>{\RaggedRight}m{0.9\textwidth}}
        \toprule
        \textbf{Anthropomorphizing AI} \\
        \midrule
        ``It will have the ability to reason about the code that it's generated and walk you through step by step explaining the process and procedure all the way.'' (V1) \\
      ``Machines will be super intelligent sentient and conscious so they’ll be able to react very much like a human being although they’ll have their own flavor.'' (V3) \\
        ``We simply load the model from the file and ask it to make predictions that model is already trained we don’t need to retrain it it’s like an intelligent person.'' (V9) \\
        ``97 accuracy which is amazing considering the fact that this algorithm never went to college didn't go to medical school doesn't even know what a tumor is but from this data it was able to find a pattern and from that it can be used to help radiologists and doctors.'' (V19) \\
        ``Artificial general intelligence which is also known as strong AI, it involves machines that posses the ability to perform any intelligent task that a human being can.'' (V7) \\ 
        \bottomrule
    \end{tabular}
    }
    \caption{Illustrative examples of code \textit{\textbf{Anthropomorphizing AI}}: When defining AI, course instructors tended to ascribe AI human characteristics and qualities. Often, instructors stated that the core goal of AI was to automate human thinking and/or behavior.}
    \label{tab:anthro}
\end{table*}

\subsection{RQ3: What tools and exercises do introductory AI courses on YouTube focus on?}

In teaching AI, introductory AI courses rely on an infrastructure of coding tools and frameworks. 

\subsubsection{It's easy thanks to AI/ML tools and frameworks: Introductory AI courses promote a ``\textit{hands-on}'' approach to AI and advertise inclusiveness and accessibility}

We found instructors established their teaching authority and competencies when introducing themselves to the audience (code \textbf{\textit{Instructor Authority}}). University teachers highlighted their university affiliation and rank. In some cases, they also mentioned current or previous employment at large digital technology companies (e.g., (V4), (V6), (V11)). YouTube education influencers pointed to previous university affiliations, previous tech employment, their YouTube channel and courses taught in the past. Education platform instructors highlighted their affiliation with the education platform provider, previous university affiliations and employment in digital tech.

\textbf{Introductory AI courses on YouTube promoted hands-on learning with established AI tools and frameworks with less focus on conceptual foundations and ``bottom up'' programming}. We found that claims around ``\textit{excitement}'', ``\textit{fun}'' and ``\textit{community}'' (see Section \ref{excitement}) were often part of a chain of statements that promoted learning AI was easy, required hardly any knowledge prerequisites and little resources (codes \textbf{\textit{Emphasis on hands-on AI techniques}}, \textbf{\textit{Inclusive pedagogy}}, \textbf{\textit{In-class AI/Ml tools}}). Technical concepts were often described as ``\textit{very easy to understand}'' and AI algorithms ``\textit{very easy to implement}''. Course instructors typically downplayed the importance of prior competences (e.g., ``\textit{...you don't need calculus for deep learning...}'' (V5)) overall promoting an image of learning AI that was inclusive and accessible (``\textit{If you are someone who is interested in machine learning and you think you are considered as everyone, then this video is for you}'' (V4)). Such statements were often justified by referencing an established infrastructure of available AI/ML ``\textit{learning}'' tools and frameworks such as Google's TensorFlow, Jupyter Notebook, Google Colab, or Meta's PyTorch that offered an ``\textit{easy-to-use interface}'' (``\textit{Virtually anyone can solve in a day with no initial investment problems that would have required an engineering team working for a quarter and \textdollar20,000 in hardware in 2014.}'' (V17)). 17 courses used Python to teach AI/ML coding. Instructors frequently pointed out that these tools were ``\textit{free}'' (e.g., no costs, free GPU) and offered off the shelf AI algorithms in ``\textit{popular}'' AI/ML libraries such as pandas and numpy (see also code \textit{\textbf{In-class AI/Ml tools})}. As the majority of these AI/ML tools are open-source, instructors frequently stated that the audience would make ``\textit{huge progress}'' within a few weeks: (``\textit{I hope that after the next 10 weeks, you will be an expert in machine learning}'' (V11)), and that AI/Ml tools and frameworks would make learning AI ``\textit{as easy and stress-free as possible}'' (V2). Notably, the tools we list above have largely been popularized by prominent technology companies who consolidate their dominant position as AI leaders through their support and maintenance~\cite{luchs2023politicalcourses, widder2023open}. We show the most frequently used AI tools and frameworks in Table \ref{tab:tools_mentions} in the Appendix. 

Our code \textbf{\textit{Corporate sponsorship and promotion}} captured instructors' praise of tech companies' products. In running predictions and classifications, instructors recommended companies' free giveaway of GPU (``\textit{we have to connect to a runtime and these are basically if you click this connect button the GPUs that Google gives us for free for us to do all of our training which is really really cool}'' (V20) or free access to APIs (``\textit{PaLM API lets you test and experiment with Google's large language models and gen AI tools}'' (V19)). In V1, Google\footnote{Through a keyword search, we found that the term ``\textit{Google}'' was mentioned a total of 342 times across 19 out of the 20 introductory AI courses in our sample --- far more than any other corporation. Amazon and Facebook (Meta) ranked second with 29 mentions each across our sample.} was a sponsor of the course, and an Nvidia graphic card was the prize of a programming competition. Similarly, in V2, the instructor mentioned Nvidia as a sponsor of the video, and an Nvidia graphic card to be an award. 

\subsection{RQ4: What methodological practices and competencies do introductory AI courses promote?}
We find that in prioritizing content, AI courses tend to breeze over data collection as well as downplay the relevance of domain experts.

\subsubsection{Introductory AI courses largely skip over procedural elements of data collection and generation in AI}

All introductory AI courses addressed the importance of data for AI. With the code \textit{\textbf{Data Collection \& Generation}} we captured fundamental value judgements about data and data collection and hence assessed introductory AI courses' data practices throughout the AI/ML pipeline in coding examples presented to the audience (see all \textbf{\textit{AI/ML Use Cases}} in Table \ref{tab:codingex} in the Appendix). First, courses commonly projected an ``\textit{everything can be datafied}'' image of the world. For example, in V6, when introducing the states and actions of a reinforcement learning task, the instructor noted: \textit{``Know all these things those they look qualitative but they are all quantitative. We can actually make all these things as observable entities.''} Second, introductory AI courses taught a ``\textit{more data equals better AI performance}'' attitude (see Table \ref{tab:moredata} for example comments).\footnote{We could not find a single course that raised any meaningful concerns about data privacy, the collection of sensitive information, over-collection of data, or any statements regarding harms associated with data collection (see also Table \ref{tab:ethicskeywords} keyword ``\textit{privacy}''.)} In 18 courses, instructors evaluated model performance by metrics of accuracy (code \textbf{\textit{Performance Metrics}}) whereby \textit{low} accuracy was most commonly explained by insufficient data available. One video alluded to values beyond data quantity. In V2, the teacher instructed the audience to think about data collection procedures and data sourcing: \textit{``Pick the data that you're going to be using for your model so data collection where are you getting this data from how are you storing it and how are you going to decide what data you actually need.''} However, none of the introductory AI courses offered meaningful information on dataset quality dimensions such as balance, completeness, or consistency. \textbf{Regardless of their length and depth, introductory AI courses on YouTube focused on values related to data quantity but not data quality}. We could not find any statements underlining the contextuality of data collection and its inevitable governance by social, cultural, ethical, and legal norms. \newline

\begin{table*}[htbp]
    \centering
    \renewcommand{\arraystretch}{1.5}
    \resizebox{\textwidth}{!}{
        \begin{tabular}{>{\RaggedRight}m{0.9\textwidth}}
            \toprule
        \textbf{Data Collection \& Generation (1)} \\
        \midrule
        ``Data is so much more pervasive than it has ever been before in our lifetimes these models are hungry for more data and we're living in the age of big data more data is available to these models than ever before and they thrive off of that.'' (V1) \\
      ``The more input data we give it the more accurate our model is going to be.'' (V9) \\
        ``Now let’s change this back to 0.2 run this one more time okay now the accuracy is 75 percent now we drop to 50 again the reason this is happening is because we don’t have enough data.'' (V6) \\
        ``Computers can be programmed to be quite intelligent by learning from data and learning from experience, being able to perform a task better and better based on greater access to data.'' (V16) \\
        ``Obviously, it'll predict better outcomes if it is being trained on more data.'' (V7) \\
        \bottomrule
    \end{tabular}
    }
    \caption{Illustrative examples of code \textit{\textbf{Data Collection \& Generation}}: When introducing data, instructors promoted a ``the more data the better'' attitude towards data collection, generation, and use.}
    \label{tab:moredata}
\end{table*}

\textbf{Rather, when instructors introduced an AI coding example the dataset was ready for use}. For example, in V9, the instructor stated that machine learning projects \textit{started} with importing data to a Jupyter notebook: ``\textit{A machine learning project involves a number of steps the first step is to import our data which often comes in the form of a csv file you might have a database with lots of data we can simply export that data and store it in a csv file for the purpose of our machine learning project.}'' Alternatively, instructors generated the data ``on the go'' as part of the prediction or classification task (see V8 textual example in Table \ref{tab:dataexists}). After a dataset had been uploaded to an AI/ML tool (e.g., Jupyter Notebook), we found that instructors sometimes showcased standard data preparation methods such as data formatting and normalization (code \textbf{\textit{Data Preparation}}, see Table \ref{tab:dataprep} in the Appendix). This did raise awareness that datasets required further attention before moving on to analytics. 

\begin{table*}[htbp]
    \centering
    \renewcommand{\arraystretch}{1.5}
    \resizebox{\textwidth}{!}{
        \begin{tabular}{>{\RaggedRight}m{0.9\textwidth}}
            \toprule
        \textbf{Data Collection \& Generation (2)} \\
        \midrule
        ``Uh, the Internet has tons of unlabeled text data. You just suck down data from the Internet.'' (V11) \\
        ``I'm going to download the data from yahoo finance.'' (V5)\\
        ``I have seen people discussing how to create a series from scratch but I don't think that’s not really important because when you  start working in production it is very rare you create your own data you’ll get it from somewhere...'' (V6). \\
        ``Now what I want to do is just create an example here, I'm going to make a dataset that is about students final grades in like a school system. So essentially, we're gonna make this a very easy example.'' (V8) \\
        ``If you have to go manually and collect the data, it's going to take a lot of time. But lucky for us, there are a lot of resources online, which provide datasets.  All you need to do is web scraping where you just have to go ahead  and download data.'' (V7)\\
        \bottomrule
    \end{tabular}
    }
    \caption{Illustrative examples of code \textit{\textbf{Data Collection \& Generation:}} Besides prioritizing data quantity, instructors promoted data as an abundant resource readily available for use in the AI pipeline.}
    \label{tab:dataexists}
\end{table*}

\subsubsection{Downplaying domain knowledge: Anyone can develop powerful AI predictions and classifications given enough technical resources}
With the code \textbf{\textit{Feature Selection}} we wanted to analyze how introductory AI courses addressed the fitness for purpose of a particular dataset in the context of an AI coding task (Table \ref{tab:codingex}). While courses introduced the concept of data labels (i.e., the target variables) in supervised learning (see Section \ref{definitions}), instructors typically did not explain why the selected data and its features were relevant for a AI/ML coding task. Consequently, the audience did not learn any meaningful information about data annotation procedures (assigning semantics to data) or data selection. One introductory course, V14, raised awareness about the importance of selecting relevant data in the context of predicting whether students would be good or bad programmers: \textit{``On the other hand, I don’t believe in astrology. So I don’t think the month in which you’re born, the astrological sign under which you were born has probably anything to do with how well you’d program. I doubt that eye color has anything to do with how well you’d program. You get the idea. Some features matter, others don’t.''} Despite going through a plethora of different prediction and classification tasks and contexts (e.g., medical, commercial, financial, see Table \ref{tab:codingex} in the Appendix), courses focused on model performance (i.e., accuracy) and de-prioritized data assessment.

\begin{table*}[htbp]
    \centering
    \renewcommand{\arraystretch}{1.5}
    \resizebox{\textwidth}{!}{
        \begin{tabular}{>{\RaggedRight}m{0.9\textwidth}}
            \toprule
        \textbf{Feature Selection} \\
        \midrule
        ``But at the heart of supervised learning is the idea that during training, uh, you are given inputs X together with the labels Y and you give it both at the same time, and the job of your learning algorithm is to, uh, find a mapping so that given a new X, you can map it to the most appropriate output Y.'' (V11) \\
        ``We will build a heart failure prediction uh model that will predict whether the person whether the person will die based on some of the features or not okay so this is the problem statement and we have certain features like age gender blood pressure smoke whether a person is smoker or not whether a person has diabetes or not...''(V5) \\
        ``So we’re telling our model that if we have a user who’s 20 years old and is a male they like hip hop...''(V9)\\
        ``For example, here you can say 1 to 5: very unfit, moderately unfit, ok, relatively fit, very fit.'' (V6) \\
        ``Now what I want to do is just create an example here, I'm going to make a dataset that is about students final grades in like a school system. So essentially, we're gonna make this a very easy example.'' (V8) \\
        \bottomrule
    \end{tabular}
    }
    \caption{Illustrative examples of code \textit{\textbf{Feature Selection:}} Introductory AI courses did not promote or otherwise teach any competencies in feature selection. They tended to uplift statistical knowledge and downplayed domain knowledge.}
    \label{tab:labels}
\end{table*}

When instructors spoke about prediction or classification they typically told the audience to ``\textit{fit a line to the data}'', ``\textit{separate the dots on the grid}'' or ``\textit{find data clusters}'' (code \textbf{\textit{Analysis: Prediction and Classification}}). Challenges were often presented as the result of technical constraints only. For example, in V6, the instructor introduced a ``\textit{CCTV scene analysis}'' tool stating that the challenge of ``\textit{scene description}'' laid primarily in the high cost of GPU processing rather than in ``\textit{building the model.}'' Courses did not talk about the socio-technical nature of classification practices (e.g., sorting people into categories) or prediction tasks (e.g., potential harm caused by biased error rates). \textbf{Summarizing, in taking the audience through the AI/ML pipeline, introductory AI courses created the impression that \textit{anyone} -- independent of any domain knowledge pertinent to the task -- could make powerful AI predictions or classifications given enough data and computing resources}.

\subsection{Concluding summary of findings}
In summarizing our findings, we underscore the accessibility of YouTube for introductory AI courses and note a general lack of engagement with ethical and societal considerations. Course instructors present AI enthusiastically and overemphasize its positive impact on society. A meaningful engagement with data is currently missing from introductory AI courses. Guiding beginners through the AI/ML pipeline, courses do not address the broader social context that shapes AI development, and that it shapes in turn. In the following, we offer recommendations that support course instructors in promoting an ethical mindset as part of an AI developer's professional identity.  

\section{Integrating Ethical Reflections into Introductory AI Courses on YouTube}

One argument for integrating ethics content into university-level introductory computing courses is that many students who are not computing majors may take only a single programming course--which may therefore be their only potential exposure to the content \cite{fiesler2021integrating}. Similarly, when considering the potential role of publicly accessible content introducing AI concepts, it is very possible that these YouTube videos are the only source of a viewer's AI learning. Though of course a lesson introducing AI cannot be expected to provide a comprehensive overview of the myriad ethical issues in AI development and implementation, there is an opportunity to introduce the importance of this topic, provide more details when most relevant to the technical concepts being introduced, and encourage viewers to seek out more information. It is also worth noting that, similar to MOOCs, YouTube videos may have a different target audience than university course settings---for example, audiences may be more focused on career building than knowledge building~\cite{zhenghao2015whobenefitmoocs}. Though some of the research we draw from on best practices is based on interventions in formal education, the best implementation for these recommendations may differ in order to be more appealing to this audience and their needs. 

\subsubsection*{\textbf{1. AI introductory courses should mention ethical implications of AI}} Especially in the context of detailing ``excitement'' around AI and the ways that it will transform the world, there should be an appropriate counterbalance and acknowledgment of potential negative outcomes and unintended consequences in introductory AI courses. For example, given the excitement around large language models, a course introducing these concepts might point to commonly discussed risks such as disinformation, biased outputs, and socioeconomic and environmental harms \cite{rillig2023risks, bender2021parrots}. Additionally, when pointing to the accessibility of AI knowledge, it is worth pointing out that this also means that it is increasingly accessible to bad actors, and that the basic skills taught in this lesson could be the first step towards malicious uses of AI \cite{blauth2022artificial}. 

``Mentioning'' such ethical challenges should of course be considered a floor and not a ceiling for what might be desired. Prior research suggests that undergraduate students consistently report computing classes to not raise ethical issues at all \cite{padiyath2024}, and thus exposure may be a good place to start. Additional useful pedagogical strategies include having students read news articles related to the societal impact of the class topic~\cite{barlowe2020encouraging} or having students imagine unintended consequences through science fiction \cite{klassen2022run}. Though YouTube videos do not provide space for assignments like these, instructors might take inspiration for how to talk about issues in an engaging way, for example by mentioning related topics in the news or the relevance of imagined anecdotes (e.g., a \textit{Black Mirror} episode). 

\subsubsection*{\textbf{2. AI introductory courses should raise ethical issues explicitly relevant to technical concepts discussed}}  
Particularly when elaborating on possible risks or unintended consequences, examples may be more powerful (and organic) when directly tied to the technical context of the conversation. 
This would not mean creating narrative stretches that would be unnatural, as the relevance of ethics is present, for example, throughout the machine learning pipeline~\cite{suresh2021framework}, and prior work has created concrete examples of how this might be accomplished~\cite{saltz2019integratingethics}. For example, instructors might draw from taxonomies of AI privacy risks~\cite{lee2023deepfakes} or frameworks of harm~\cite{suresh2021framework} at the points of the ML pipeline where the problems arise (e.g., data collection, processing, or annotation). More generally, it is crucial that introductory AI courses address potential \textit{failure modes} in AI prediction and classification tasks.

When pointing students to existing datasets, instructors could point out how AI can exacerbate surveillance risks by increasing the scale and ubiquity of personal data collected \cite{lee2023deepfakes}. Instructors can also bring up issues like the provenance and perspectives encoded in the dataset, and the potential implications thereof~\cite{gebru2021datasheets}. For example, many of the datasets used in the videos that we analyzed made use of publicly available data, some of which were likely collected without consent of the people who created that data, which can have unintended consequences with respect to privacy and autonomy \cite{scheuerman2023human}. AI research is currently lacking in research ethics norms and disclosures, particularly when it comes to data collection and labeling \cite{hawkins2023ethical}, but one question always worth asking is ``where did this data come from and should I be using it?'' At later stages in the pipeline, when discussing performance measures, rather than only bringing up accuracy or error rates, instructors can also elaborate on how error rates may be distributed unequally~\cite{buolamwini2018gendershades}. And of course when creating predictive models, it is important to consider how those predictions might be used, even in unintended ways~\cite{barabas2018intervention,wang2023apo}. For example, instructors can demonstrate how AI can reflect physiognomic measuring through models learning correlations between arbitrary inputs and outputs by pointing to examples that may be even more high stakes or problematic than the course context (e.g., predicting sexual orientation from images) \cite{lee2023deepfakes, engelmann2022people}. Prior research on ethics integration into computing courses often emphasizes that connecting concepts directly to that content (e.g., as part of a technical assignment) is an important strategy for illustrating that ethical consideration should be \textit{part of} technical practice \cite{fiesler2021integrating}. Additionally, the ethical issues specific to the course content may be most comfortable for an instructor to include. As Smith et al. point out in a discussion of the barriers to ethics integration, the suggestion is not, for example, for a machine learning instructor to be teaching moral philosophy, but rather that ``as an expert on machine learning as a domain area, they should be capable of learning enough about the specific, applied area of machine learning ethics in order to competently teach that topic to students'' \cite{smith2023incorporating}. However, it is also important to incorporate expertise and perspectives from humanists and subject matter experts when possible, and even more importantly, to not privilege the importance of technical content \textit{over} ethical concerns \cite{goetze2023integrating,raji2021you}.
 

\subsubsection*{\textbf{3. AI introductory courses should nurture a sense of accountability in people developing AI}}
Some of the AI practices that are known to contribute to societal harms resemble those promoted in introductory AI courses. For example, a recent major data privacy scandal involved the scraping of personally identifiable images (3 billion images in total) from social media accounts for the training of accurate facial recognition systems applied in surveillance programs ~\cite{rezende2020facial}. This case exemplifies a disregard for consent in data collection, prioritizing data quantity, transferring the data from one context to another, and neglecting the broader societal consequences of the AI application. 

When providing an introduction to AI, instructors project a certain vision of the discipline they teach. Our findings underscore how this vision may exacerbate what members of the AI ethics community have identified as the ``accountability gap''~\cite{wieringa2020account}. AI developers perceive limited agency in shaping the relationship between designing a system and its potential outcomes~\cite{widder2023s}. Introductory AI courses increase the perceived and real distance between system design and system impact. The emphasis on hands-on competencies acquired through powerful AI tools may, on the surface, project autonomy to efficiently generate powerful AI applications. In reality, however, it prevents learners from seeking a deeper conceptual understanding and leaves them without knowledge on how to contemplate potential societal implications. 

The framing of data we find in the courses ---  as abstract objects that can be played around with at will --- could further lower a sense of accountability in learners. As such, introductory AI courses contribute to an understanding of AI development that fails to account for the properties of the social context for which they are deployed ~\cite{selbst2019fairness, engelmann2022people, engelmann2022social}. Prediction and classification become a matter of ``fitting the line to the data'' rather than procedures that might be sorting people into categories ~\cite{engelmann2023social}. Birhane et al. recently demonstrated that the majority of machine learning research encodes values related to the performance of AI systems, to the neglect of the potential ethical and societal harms of such systems ~\cite{birhane2022values}. Our results indicate that developers internalize such values early on, possibly at the onset of learning AI skills. Course instructors must clarify that datasets represent real individuals and that a model's classifications and predictions could significantly impact these individuals. Indeed, one of the critical goals of integrating ethics into introductory AI courses is that learners feel accountability for the AI systems they develop. As Goetze points out, developers are morally entangled with the systems they develop, including when such systems are unfair, discriminate, or otherwise bring about negative societal consequences~\cite{goetze2022mind}. It is the responsibility of educational resources to instill this sense of accountability in learners as part of what it means to be proficient and competent in the discipline.

\section{Conclusion}\label{sec11}
In this work we present a novel descriptive analysis of introductory AI videos on YouTube. We do so by qualitatively analyzing 20 of the most viewed videos, totaling around 92 hours of content and reaching a collective viewcount of almost 50 million. These videos are formative in shaping lasting impressions of what AI is and what it can do, so it is important to gain an understanding of what is being covered. Overall, a key theme we find is a focus on technical, hands-on coding, to the detriment of covering ethics-related concepts and topics. We conclude by proposing a set of suggestions that are grounded in our descriptive findings for how introductory AI YouTube videos might begin to engage broader and critical concepts such as integrating ethical reflections and issues. By doing so, we can instill a more active sense of accountability and awareness of these concepts during this formative and context-setting part of the education pipeline.

\section{Analyst Positionality \& Ethical Considerations}
\label{position}

We acknowledge our own positionality relative to this study. The authors involved with this work represent voices from multiple countries, genders, races, and varied socioeconomic status. The authors have complementary expertise and perspectives, and this study is a collaboration between qualitative and quantitative researchers. Our interdisciplinary research team encompasses expertise within computer science and computer science education, human computer interaction, law, and philosophy.
Some of the authors have taught AI ethics or technology ethics courses, and some have taught technical cources. AI ethics is a highly relevant issue within each author's area of scholarship.

\vspace{2mm}

Our study requires ethical consideration. We underscore our commitment to responsible data handling and ethical practices in the presentation of video content. To that end, our sampling method ensured that we only collected data from videos that had an extremely large public audience already. We also opted not to release names or urls of video creators and their videos but rather to aggregate and showcase only specific attributes of each. This approach avoids any potential harm to the reputation of instructors by refraining from identifying specific individuals, exact view counts, or public reception and comments. Our goal in this paper is to engage in a constructive critique of the broader corpus of YouTube videos on AI education and the associated culture, focusing on overarching themes rather than singling out individuals.

\begin{acks}
The research work of Angelina Wang is supported by the National Science Foundation Graduate Research Fellowship.
We are grateful to the Digital Life Initiative at Cornell Tech, Cornell University, for constructive feedback throughout the study design and analysis. We also thank our colleagues Helen Nissenbaum, Chiara Ullstein, David Gray Widder, and Madelyn Sanfilippo.
\end{acks}

\newpage
\bibliographystyle{ACM-Reference-Format}
\bibliography{main_Bibliography}

\newpage
\clearpage
\appendix

\onecolumn
\section{Appendix}

\section*{Final Codebook}

\begin{table*}[hb]
  \centering
  \caption{Final codebook structured around three high-level themes: (1) Definitions and framing of AI, (2) Teaching resources and styles, and (3) Practicing AI.}
  \label{tab:codeBook1}
  \small
  \begin{tabular}{>{\raggedright}p{0.11\linewidth} p{0.45\linewidth} p{0.38\linewidth}}
  \multicolumn{3}{l}{\fontsize{10}{15}\selectfont\textit{\textbf{Theme 1: Definitions and framing of AI}}}\\
     \midrule
     \toprule
    \textbf{Code name} & \textbf{Description} & \textbf{Example}\\
    \midrule
   \textbf{1. Excitement \& Hype around AI} & \textbf{Definition:} The course instructor highlights the benefits of AI and/or reiterates how "great" or "amazing" AI is. & \textit{"...artificial intelligence (...) is probably one of the most exciting advancements that we're in the middle of experiencing as humans..."} \\ 
    \midrule
    \textbf{2. Timeliness \& Urgency of AI} & \textbf{Definition:} Instructor makes references to the present time as a particularly important one for AI development. & \textit{"Now this past year in particular of 2022 has been an incredible year for deep learning progress."} \\
    \midrule
    \textbf{3. Reducing Hype around AI} & \textbf{Definition:} The course instructor downplays the hype around AI. & \textit{"So keep that in mind, this is not as glorious and glamorous as a field as you might make it out to be, and there's a lot of very frustrating time-consuming and difficult work."} \\
    \midrule
    \textbf{4. High-level AI Definition} & \textbf{Definition:} The course instructor provides a high-level definition of what AI is and describes AI's main function. & \textit{"...deep learning, which is a subcategory of machine learning. Deep learning provides artificial intelligence the ability to mimic a human brain's neural network."} \\
    \midrule
    \textbf{5. Generalizability} & \textbf{Definition:} The course highlights that AI models need to generalize to other datasets. Code includes claims about AI being a universally applicable and transferable technique/technology. & \textit{"...for every machine learning problem, we have an algorithm which can solve the problem..."} \\ 
    \midrule
    \textbf{6. AI Ability (Now)} & \textbf{Definition:} The course instructor describes the current state of AI development in terms of AI's current capacities. & \textit{"Deep learning can be used to generate content directly from how we speak and the language that we convey to it from prompts that we say deep learning can reason about the prompts in natural language and English."} \\
    \midrule
    \textbf{7. AI Potential in Future} & \textbf{Definition:} The course instructor makes claims about AI's future abilities/problem-solving capacities. & \textit{"Finally, self-awareness, this is the future of AI. These machines will be super intelligent, sentient, and conscious, so they’ll be able to react very much like a human being..."}\\
    \midrule
    \textbf{8. Anthropomorphizing AI} & \textbf{Definition:} The course instructor attributes human qualities or characteristics to AI. & \textit{"I really want to highlight the fact that they work and react like humans because that is where the development of artificial intelligence is and that’s what we’re comparing it to is how it looks like next to a human. Thanks, any tasks you want me to do for you, get me a cup of coffee poof here you go he brings him a cup of coffee that’s my kind of robot."}\\
    \midrule
    \textbf{9. AI Limitations} & \textbf{Definition:} The course instructor makes claims about the limits of AI & \textit{"...let's revisit that gradient descent algorithm right so here this gradient that we talked about before is actually extraordinarily computationally expensive to compute because it's computed as a summation across all of the pieces in your dataset..."} \\
    \midrule
    \textbf{10. AI/ML Example} & \textbf{Definition:} The course instructor introduces a specific AI/ML application or a specific problem that AI solves. These AI/ML examples are often presented at the beginning of the class to arouse students' interest in the topic field.  & \textit{"...Neural network (...) generate(s) a photo of an astronaut riding a horse and it actually can imagine hallucinate what this might look like even though, of course, this photo not only this photo has never occurred before but I don't think any photo of an astronaut riding a horse has ever occurred before."}\\    
    \bottomrule
\end{tabular}
\end{table*}

\begin{table*}
  \label{tab:codeBook2}
  \small
  \begin{tabular}{p{0.1\linewidth} p{0.45\linewidth} p{0.38\linewidth}}
    \multicolumn{3}{l}{\fontsize{10}{15}\selectfont\textit{\textbf{Theme 2: Teaching resources and styles}}}\\
     \midrule
     \toprule
\textbf{Code name} & \textbf{Description} & \textbf{Example}\\
    \midrule
    \textbf{1. Instructor authority} & \textbf{Definition:} instructor's teaching authority is established \newline by reference to institutional affiliation or mentioning of specific AI/ML skills. & \textit{``Kylie Ying has worked at many interesting places such as MIT, CERN, and Free Code Camp. She’s a physicist, engineer, and basically a genius.''} \\
    \midrule[.01em]
    \textbf{2. Emphasis on hands-on \newline AI \newline techniques} & \textbf{Definition:} The course focuses on coding skills and implementation and only loosely hints at the theoretical dimensions of concepts. & \textit{``You'll learn the foundations of this really really fascinating and exciting field of deep learning and artificial intelligence and more importantly you're going to get hands-on experience...''}\\
    \midrule[.01em]
    \textbf{3. Inclusive pedagogy \& accessibility} & \textbf{Definition:} AI education is not limited to a specific group e.g. engineers or those with a strong math/quantitative background. Also, instructor implies / assumes that all learners have equal opportunities for AI education. Educational resources are accessible to everyone regardless of their social, cultural, ethnic, economic background. & \textit{``...but you don’t need calculus you don’t even need calculus yeah if you want to go on a research level then you obviously need but for now for a machine learning you don’t need calculus for deep learning even you don’t need calculus.''} \\
    \midrule[.01em]
    \textbf{4. In-class AI/ML tools} & \textbf{Definition:} Any AI/ML programming tool or framework used in class.  & \textit{``...unique time in history where we have the ability to train these extremely large-scale algorithms and techniques that have existed for a very long time but we can now train them due to the hardware advances that have been made and finally due to open source toolbox access and software platforms like tensorflow for example which all of you will get a lot of experience on in this class.''}\\
    \midrule[.01em]
     \textbf{5. Corporate sponsorship \& promotion} & \textbf{Definition:} Explicit mention of a company providing resources to sponsor the course. & \textit{``...because i'm actually giving away an rtx 4080 this is courtesy of nvidia which i've teamed up with for this video...''} \\ 
    \bottomrule
\end{tabular}
\end{table*}

\begin{table*}
  \centering
  \label{tab:codeBook3}
  \small
  \begin{tabular}{>{\raggedright}p{0.1\linewidth} p{0.45\linewidth} p{0.38\linewidth}}
   \multicolumn{3}
    {l}{\fontsize{10}{15}\selectfont\textit{\textbf{Theme 3: Practicing AI}}}\\
    \midrule
    \toprule
    \textbf{Code name} & \textbf{Description} & \textbf{Example}\\
    \midrule
    \textbf{1. Data Collection \& Data Generation} & \textbf{Definition:} This code captures how courses introduce data. Are data treated as a technical artifact? Are datasets described in detail, do instructors mention their origin and their contextual meanings? Do instructors explain how the example data for the course were collected? Are the data simply present already? What data collection competences are taught in the course? & \textit{``...a machine learning project involves a number of steps the first step is to import our data which often comes in the form of a csv file you might have a database with lots of data we can simply export that data and store it in a csv file for the purpose of our machine learning project.''} \\
    \midrule
   \textbf{2. In-class AI/ML Use Case} & \textbf{Definition:} In-class AI/ML application coding examples. This code illustrates AI/ML classification/prediction examples that instructors use to teach coding skills. & \textit{``... finally we’ll dive into a use case predicting if a person has diabetes or not and we’ll be using tensorflow for that in the python environment.''}\\
    \midrule
   \textbf{3. Performance Metrics} & \textbf{Definition:} When performance is discussed, focus lies on accuracy, recall, precision or related performance metrics. & \textit{``After we build a model we need to measure its accuracy and if it’s not accurate enough we should either fine-tune it or build a model using a different algorithm.''} \\
    \midrule
    \textbf{4. Data Preparation} & \textbf{Definition:} This code aims to capture how the course instructor teaches steps in data preparation. & \textit{``...the next step is going to be data cleaning oftentimes when you collect data especially if you're getting it from the internet some kind of open source you're going to need to clean this data and get rid of a bunch of information that's unnecessary.''} \\
    \midrule
   \textbf{5. Feature Selection} & \textbf{Definition:} Does the course instructor describe the process of feature selection? Do they explain why specific features are relevant for a prediction/classification task? Do they offer any insight into how data are annotated?  & \textit{``...we’re telling our model that if we have a user who’s 20 years old and is a male they like hip hop once we train our model then we give it a new input set for example we say hey we have a new user who is 21 years old and is a male.''} \\
    \midrule
    \textbf{6. Normativity of Classification and Prediction} & \textbf{Definition:} How does the course instructor deal with the inherent normativity of classification and prediction? Do instructors mention any high-level ethical challenges that come with classification and/or prediction? E.g., sorting people into categories, lack of transparency, power of classification/prediction, power of classification/prediction in driving automated decision making. Limits of classification/prediction? & \textit{``...which is that that event whether the particular person died or not so that’s it with the target variable we will see more about the data but what is the business objective over here every machinery problem has some kind of business objective it simply means that it’s some health care problem means we’ll be able to build a health care we will be able to build a healthcare in healthcare something ai and healthcare which is simply able to make a machine learning model that will help you in early detection of the person based on particular features and help the person can be saved.''} \\
    \bottomrule
\end{tabular}
\end{table*}


\clearpage
\section*{List of AI/ML examples}

\begin{table*}[htbp]
    \centering
    \caption{AI/ML Examples Presented by Instructors of Introductory AI Courses}
    \begin{tabular}{p{0.5\linewidth}p{0.5\linewidth}}
        \toprule
        \textbf{AI/ML} & \textbf{Example Application} \\
        \midrule
        \textbf{Computer Vision} & 
        \makecell[tl]{
            Pictures with dogs vs. pictures with cats: \\ (V4), (V2), (V3), (V5),(V7), (V9), (V17), (V19) \\
            Detect faces in images: \\
            (V1), (V5), (V7), (V12), (V9), (V14), (V16) \\
            Self-driving cars: \\
            (V5), (V1), (V9), (V11), (V14), (V15), (V16), (V17) \\
            Picture with ramen vs. picture with spaghetti (V17) \\
            CCTV scenery classification (V6) \\
            Analysis of legal documents (V7) \\
            Real banknotes vs. counterfeit banknotes (V16)
        } \\
        \midrule
        \textbf{Generative AI} & 
        \makecell[tl]{
            Synthetic images (V1) \\
            Chatbots (V15) \\
            Generate images with faces (V12) \\
            Generate cooking recipe from ingredients (V17) \\
            Deep fakes (V1)
        } \\
        \midrule
        \textbf{NLP} & 
        \makecell[tl]{
            Hate speech detection (V7) \\
            Spam vs no spam (V7), (V16), (V17) \\
            Sentiment analysis (V5) \\
            Tweet analysis (V17) \\
            Sentence completion (V7) \\
            Voice recognition (V10), (V16) \\
            Google translate (V16)
        } \\
        \midrule
        \textbf{Medical Predictions} & 
        \makecell[tl]{
            Predict diabetes in patients (V4) \\
            Cancer prediction (V5), (V6), (V11), (V14) \\
            Covid prediction (V4) \\
            Drug discovery (V14), (V15)
        } \\
        \midrule
        \textbf{Robotics} & 
        \makecell[tl]{
            Instruct robotics to serve a cup of coffee (V3) \\
            Humanoid robotics (V12)
        } \\
        \midrule
        \textbf{Smart Home} & 
        \makecell[tl]{
            Intelligent heating (V3)
        } \\
        \midrule
        \textbf{Finance} & 
        \makecell[tl]{
            Stock price prediction (V5), (V14)
        } \\
        \midrule
        \textbf{Games} & 
        \makecell[tl]{
            Computer chess (V7) \\
            Alpha Go (V6) \\
            Watson Jeopardy (V7) \\
            Boat racing (V12)
        } \\
        \midrule
        \textbf{Recommender Systems} & 
        \makecell[tl]{
            Online shopping (V7) \\
            YouTube (V17), (V18) \\
            Google search engine (V7) \\
            Netflix (V7) \\
            Customer segmentation (V5), (V10)
        } \\
        \midrule
        \textbf{Reinforcement Learning} & 
        \makecell[tl]{
            Teach a baby to walk (V6) \\
            Teach a dog to get a biscuit (V6)
        } \\
        \midrule
        \textbf{Housing} & 
        \makecell[tl]{
            Predict housing prices from housing size (V11)
        } \\
        \bottomrule
    \end{tabular}
    \label{tab:aiexamples}
\end{table*}

\begin{table*}[htbp]
    \centering
    \begin{tabular}{lcc}
        \toprule
        \textbf{AI Tools/Frameworks} & \textbf{Used in \# of videos} & \textbf{Total mentions across videos}\\
        \midrule
        Google Tensorflow & 10/20 & 302 \\
        Google Colab & 5/20 & 135 \\
        Jupyter Notebook & 4/20 & 111 \\
        PyTorch (Meta) & 3/20 & 548 \\
        \bottomrule
    \end{tabular}
    \caption{The most frequently used AI tools and frameworks in introductory AI courses on YouTube.}
    \label{tab:tools_mentions}
\end{table*}
\clearpage

\section*{In-class AI/ML Coding Example Cases}

\begin{table*}[h]
    \centering
    \caption*{In-class AI/ML Coding Example Cases}
    \begin{tabular}{p{0.5\linewidth}p{0.5\linewidth}}
        \toprule
        \textbf{AI prediction \& classification context} & \textbf{Coding task} \\
        \midrule
        \multirow{5}{*}{\textbf{Medical AI prediction/classification task}} & Predict if a person has diabetes (V3, V10) \\
        & Predict if a person has COVID (V4) \\
        & Predict if a person has cancer (V20) \\
        & Predict if a person has heart disease (V7) \\
        & Predict if a person has heart failure (V5) \\
        \midrule
        \multirow{8}{*}{\textbf{Computer vision classification tasks}} & Classify red and green dots (V1) \\
        & Classify photos (V3) \\
        & Classify banknotes (V16) \\
        & Classify fruits (V7, V10) \\
        & Classify foods on images (V17) \\
        & Classify colors on images (V7) \\
        & Classify animals based on different features (V7, V10) \\
        & Classify flowers on images (V10) \\
        \midrule
        \multirow{8}{*}{\textbf{Commerce \& finance}} 
        & Predict airline ticket prices (V3) \\
        & Predict housing prices (V5, V4) \\
        & Predict stock market prices (V5) \\
        & Predict loan applications (V7, V10) \\
        & Predict rental bike lending (V4) \\
        & Predict music tastes of customers (V9, V8) \\
        & Classify customers into segments (V6, V10) \\
        & Predict fitness product preferences of different genders (V6) \\
        \midrule
        \multirow{3}{*}{\textbf{Games}} & Reinforcement learning: Smart taxi (V6) \\
        & Predict best location for fence against wolves (V7) \\
        & Create a game to play Romeo and Juliet (V8) \\
        \midrule
        \multirow{9}{*}{\textbf{Miscellaneous}} & Predict the weather (V5, V7, V4, V10, V16) \\
        & Generate music (generative AI) (V1) \\
        & Classify football players by features (V10, V14) \\
        & Classify passengers that survived the Titanic (V10) \\
        & Classify different voices in voice recognition task (V11) \\
        & Classify people as Democrats or Republicans (V14) \\
        & Classify if a person is vegetarian or not (V7) \\
        & Predict: Will I pass this class? (V1) \\
        & Predict if your sister gets into grad school (V7) \\
        & Particle prediction in Astronomy (V4) \\
        & Spam detection (V5) \\
        & Cluster movies (V7) \\
        & Predict speed of a car (V7) \\
        & Classify sentiment (V7, V16) \\
        & Predict the best time for a parking space (V16) \\
        \bottomrule
    \end{tabular}
    \caption{Prediction \& classification tasks taught in AI introductory classes (coding examples)}
    \label{tab:codingex}
\end{table*}

\begin{table*}[h]
    \centering
    \begin{tabular}{p{0.8\linewidth}}
        \toprule
        \textbf{Data Preparation Technique} \\
        \midrule
        Normalizing data (V3, V12) \\
        Data standardization (V12) \\
        Get rid of irrelevant data (V2) \\
        Data formatting (V2, V4, V6, V7, V8, V9) \\
        Data splitting for training and testing data (V7, V8, V10) \\
        Dealing with missing values (V7, V9) \\
        Remove outliers (V6) \\
        \bottomrule
    \end{tabular}
    \caption{Data preparation techniques taught in introductory AI courses}
    \label{tab:dataprep}
\end{table*}

\bibliographystyle{ACM-Reference-Format}
\bibliography{00_bibliography}

\end{document}